\documentclass[aps,twocolumn,showpacs]{revtex4}
\usepackage{amsmath}
\usepackage{epsfig}
\usepackage{booktabs} 
\usepackage{multirow}

\begin{document}

\title{Investigation of meson-meson interaction}

\author{Yuheng Wu$^1$}\email[E-mail: ]{191002007@njnu.edu.cn}
\author{Xin Jin$^1$}\email[E-mail: ]{181002005@njnu.edu.cn}
\author{Hongxia Huang$^1$}\email[E-mail: ]{hxhuang@njnu.edu.cn (Corresponding author)}
\author{Jialun Ping$^1$}\email[E-mail: ]{jlping@njnu.edu.cn (Corresponding author)}
\author{Xinmei Zhu$^2$}\email[E-mail: ]{zxm_yz@126.com}
\affiliation{$^1$Department of Physics, Nanjing Normal University, Nanjing, Jiangsu 210097, China}
\affiliation{$^2$Department of Physics, Yangzhou University, Yangzhou 225009, P. R. China}

\begin{abstract}
In the framework of the quark delocalization color screening model, we investigate the meson-meson interaction in the fully light four-quark system . The calculation of the effective potentials of all the $S-$wave states shows that for most states the interaction between two vector mesons is attractive; the one between a pseudoscalar meson and a vector meson is repulsive or weakly attractive; and the one between two pseudoscalar mesons is always repulsive. However, there is still some exception. The interaction of the $IJ=00$ $\pi\pi$ channel is attractive, while the one of the $IJ=02$ $\phi\phi$ channel is repulsive. So it is difficult to use the $S-$wave $\phi\phi$ state to explain the $X(2239)$ at present calculation. The $S-$wave $\rho\rho$ states are more likely to be resonance states, which are worthy of investigating in future work. The study of the contribution of each interaction term shows that both the one-gluon exchange and the kinetic energy interaction are important in the interaction between two mesons. The study of the variation of the delocalization parameter indicates that the contribution of the kinetic energy relates to the intermediate-range attraction mechanism in QDCSM, which is achieved by the quark delocalization.
\end{abstract}


\maketitle

\setcounter{totalnumber}{5}

\section{\label{sec:introduction}Introduction}
The study of the hadron-hadron interaction is one of the most active frontiers. The study of nucleon-nucleon ($NN$) interaction has
lasted over seventy years. The quantitative description of $NN$ interaction has been achieved in the one-boson-exchange models, the chiral perturbation theory and quark models. The study of $\pi\pi$ interaction is also a classical subject in the field of strong interactions. The $\pi\pi$ scattering process has been studied as an important test of the strong interaction.

With the increasing experimental information of the meson spectrum, it becomes more and more important to develop a consistent understanding of the observed mesons from a theoretical point of view. For the low-lying vector and pseudoscalar mesons, this has been done quite successfully within the simple quark model by assuming the mesons to be quark-antiquark ($q\bar{q}$) states. For the scalar mesons, however, some questions still remain to be answered. One of the most noteworthy issues is the nature of the experimentally observed mesons $f_{0}(980)$ and $a_{0}(980)$. In a long-standing controversial discussion, the $f_{0}(980)$ has been described as a conventional $q\bar{q}$ meson~\cite{Morgan}, a $K\bar{K}$ molecule~\cite{KK1,KK2,KK3}, or a tetraquark state~\cite{Jaffe}.
In 2018, the BESIII collaboration reported the observation of the $a_{0}(980)-f_{0}(980)$ mixing ~\cite{BES1}, which would improve the understanding of the nature of $f_{0}(980)$ and $a_{0}(980)$.

In 2019, the BESIII collaboration analyzed the cross section of the $e^{+}e^{+}\rightarrow K^{+}K^{-}$ process at the center-of-mass energy range from 2.00 to 3.08 GeV. A resonant structure $X(2239)$ was observed, which has a mass of $2239.2\pm7.1\pm11.3$ MeV and the width of $139.8\pm12.3\pm20.6$ MeV ~\cite{BES2}. This state aroused the interest of theoretical physicists, and it has been studied extensively. Ref.~\cite{lqf} investigated the mass spectrum of the $ss\bar{s}\bar{s}$ tetraquark states within the relativized quark model, and found that the $X(2239)$ can be assigned as a $P-$wave $1^{--}$ $ss\bar{s}\bar{s}$ tetraquark state. Ref.~\cite{zjt} assigned the $X(2239)$ to be a hidden-strange molecular states from $\Lambda\bar{\Lambda}$ interaction.

Quantum chromodynamics(QCD) is the basic theory describing the strong interaction. However, the low-energy physics of QCD, such as the structure of hadrons, hadron-hadron interactions, and so on, is difficult to calculate directly from QCD, because of the nonperturbative complication. Lattice QCD has provided numerical results describing quark confinement between two static colorful quarks, a preliminary
picture of the QCD vacuum and the internal structure of hadrons in addition to a phase transition of strongly interacting
matter. But a satisfying description of multiquark system is out of reach of the present calculation. Various QCD-inspired quark models have
been developed to get physical insights into the hadron-hadron interaction and multiquark systems. There are the cloudy bag model~\cite{cloudy}, MIT bag model~\cite{MIT}, Skyrme topological soliton model~\cite{skyrme}, Friedberg-Lee non-topological soliton model~\cite{soliton}, the constituent quark model~\cite{Rujula,Isgur}, etc. Different models use quite different effective degrees of freedom, which might be indicative of the nature of low-energy QCD.

Among many phenomenological models, the quark delocalization color screening model (QDCSM) was developed in the 1990s with the aim of explaining the similarities between nuclear (hadronic clusters of quarks) and molecular forces~\cite{QDCSM0}. In this model, the intermediate-range attraction is
achieved by the quark delocalization, which is like the electron percolation in molecules. The color screening provides an
effective description of the hidden color channel coupling~\cite{Huang1}, and leads to the possibility of the quark delocalization.
The QDCSM gives a good description of $NN$ and $YN$ interactions and the properties of
deuteron~\cite{QDCSM1}. It is also employed to calculate the baryon-baryon scattering phase shifts
and predict the dibaryon candidates $d^{*}$~\cite{Ping1} and $N\Omega$~\cite{Huang2}. Besides, it has been used for a systematic search of dibaryon candidates in the $u$, $d$, and $s$ three flavor world~\cite{PRC51}, and the law of baryon-baryon interaction was proposed. In QDCSM, the interaction between two decuplet baryons is almost deeply attractive; the one between a decuplet baryon and a octet baryon is always weakly attractive; and the one between two octet baryons is mostly repulsive or weakly attractive. So it is interesting to extend this model to the study of the meson-meson interaction, which will help us to understand the low-energy physics of QCD and explore the nature of the new hadron states.

The structure of this paper is as follows. A brief introduction of the quark model and wave functions are given in section II. Section III is devoted to the numerical
results and discussions. The summary is shown in the last section.

\section{MODEL AND WAVE FUNCTIONS}

\subsection{The quark delocalization color screening model (QDCSM)}
The quark delocalization color screening model (QDCSM) has been described in detail in the refs.~\cite{QDCSM0,QDCSM1}. Here,we just present the salient
of the model. The Hamiltonian of the model is
\begin{widetext}
\begin{eqnarray}
H & = & \sum_{i=1}^4 \left( m_i+\frac{p_i^2}{2m_i}\right)-T_{CM} +\sum_{j>i=1}^4
\left(V_{ij}^{CON}+V_{ij}^{OGE}+V_{ij}^{OBE} \right),\\
V_{ij}^{CON} & = & \left \{
\begin{array}{ll}
-a_{c}\boldsymbol {\mathbf{\lambda}}^c_{i}\cdot
\boldsymbol{\mathbf{ \lambda}}^c_{j}~\left(r_{ij}^2+a^{0}_{ij}\right),&
   \mbox{if \textit{i},\textit{j} in the same baron orbit}\\
-a_{c}\boldsymbol {\mathbf{\lambda}}^c_{i}\cdot
\boldsymbol{\mathbf{\lambda}}^c_{j}~(
\frac{1-e^-\mu_{ij}\mathbf{r}_{ij}^2}{\mu_{ij}}+a^0_{ij}),& \mbox{otherwise}
\end{array}
\right.\label{QDCSM-vc}\\
V^{OGE}_{ij} & = & \frac{1}{4}\alpha_s \boldsymbol{\lambda}^{c}_i \cdot
\boldsymbol{\lambda}^{c}_j
\left[\frac{1}{r_{ij}}-\frac{\pi}{2}\delta(\boldsymbol{r}_{ij})(\frac{1}{m^2_i}+\frac{1}{m^2_j}
+\frac{4\boldsymbol{\sigma}_i\cdot\boldsymbol{\sigma}_j}{3m_im_j})-\frac{3}{4m_im_jr^3_{ij}}
S_{ij}\right] \label{sala-vG} \\
V^{OBE}_{ij} & = & V_{\pi}( \boldsymbol{r}_{ij})\sum_{a=1}^3\lambda
_{i}^{a}\cdot \lambda
_{j}^{a}+V_{K}(\boldsymbol{r}_{ij})\sum_{a=4}^7\lambda
_{i}^{a}\cdot \lambda _{j}^{a}
+V_{\eta}(\boldsymbol{r}_{ij})\left[\left(\lambda _{i}^{8}\cdot
\lambda _{j}^{8}\right)\cos\theta_P-(\lambda _{i}^{0}\cdot
\lambda_{j}^{0}) \sin\theta_P\right] \label{sala-Vchi1} \\
V_{\chi}(\boldsymbol{r}_{ij}) & = & {\frac{g_{ch}^{2}}{{4\pi
}}}{\frac{m_{\chi}^{2}}{{\
12m_{i}m_{j}}}}{\frac{\Lambda _{\chi}^{2}}{{\Lambda _{\chi}^{2}-m_{\chi}^{2}}}}%
m_{\chi} \left\{(\boldsymbol{\sigma}_{i}\cdot
\boldsymbol{\sigma}_{j})
\left[ Y(m_{\chi}\,r_{ij})-{\frac{\Lambda_{\chi}^{3}}{m_{\chi}^{3}}}%
Y(\Lambda _{\chi}\,r_{ij})\right] \right.\nonumber \\
&& \left. +\left[H(m_{\chi}
r_{ij})-\frac{\Lambda_{\chi}^3}{m_{\chi}^3}
H(\Lambda_{\chi} r_{ij})\right] S_{ij} \right\}, ~~~~~~\chi=\pi, K, \eta, \\
S_{ij}&=&\left\{ 3\frac{(\boldsymbol{\sigma}_i
\cdot\boldsymbol{r}_{ij}) (\boldsymbol{\sigma}_j\cdot
\boldsymbol{r}_{ij})}{r_{ij}^2}-\boldsymbol{\sigma}_i \cdot
\boldsymbol{\sigma}_j\right\},\\
H(x)&=&(1+3/x+3/x^{2})Y(x),~~~~~~
 Y(x) =e^{-x}/x. \label{sala-vchi2}
\end{eqnarray}
\end{widetext}
Where $S_{ij}$ is quark tensor operator; $Y(x)$ and $H(x)$ are standard
Yukawa functions; $T_c$ is the kinetic energy of the center of mass;
$\alpha_s$ is the quark-gluon coupling constant;
$g_{ch}$ is the coupling constant for chiral field,which is determined from the
$NN\pi$ coupling constant through
\begin{equation}
\frac{g_{ch}^2}{4\pi}=\left(\frac{3}{5}\right)^2\frac{g_{\pi NN}^2}{4\pi}\frac{m_{u,d}^2}{m_N^2}
\end{equation}
The other symbols in the above expressions have their usual meanings. All model parameters, which are determined by fitting the
meson spectrum, are from our previous work of tetraquark $X(2900)$~\cite{XueY}.
A phenomenological color screening confinement potential is used here, and $\mu_{ij}$ is the color screening parameter, which is determined by fitting the
deuteron properties, $NN$ scattering phase shifts, and $N\Lambda$ and $N\Sigma$ scattering phase shifts, with
$\mu_{uu}=0.45~fm^{-2}$, $\mu_{us}=0.19~fm^{-2}$, $\mu_{ss}=0.08~fm^{-2}$, satisfying the relation,
$\mu_{us}^2=\mu_{uu}\mu_{ss}$.

The quark delocalization in QDCSM is realized by specifying the
single particle orbital wave function of QDCSM as a linear
combination of left and right Gaussians, the single particle
orbital wave functions used in the ordinary quark cluster model,
\begin{eqnarray}
\psi_{\alpha}(\mathbf{s}_i ,\epsilon) & = & \left(
\phi_{\alpha}(\mathbf{s}_i)
+ \epsilon \phi_{\alpha}(-\mathbf{s}_i)\right) /N(\epsilon), \nonumber \\
\psi_{\beta}(-\mathbf{s}_i ,\epsilon) & = &
\left(\phi_{\beta}(-\mathbf{s}_i)
+ \epsilon \phi_{\beta}(\mathbf{s}_i)\right) /N(\epsilon), \nonumber \\
N(\epsilon) & = & \sqrt{1+\epsilon^2+2\epsilon e^{-s_i^2/4b^2}}. \label{1q} \\
\phi_{\alpha}(\mathbf{s}_i) & = & \left( \frac{1}{\pi b^2}
\right)^{3/4}
   e^{-\frac{1}{2b^2} (\mathbf{r}_{\alpha} - \mathbf{s}_i/2)^2} \nonumber \\
\phi_{\beta}(-\mathbf{s}_i) & = & \left( \frac{1}{\pi b^2}
\right)^{3/4}
   e^{-\frac{1}{2b^2} (\mathbf{r}_{\beta} + \mathbf{s}_i/2)^2}. \nonumber
\end{eqnarray}
Here $\mathbf{s}_i$, $i=1,2,...,n$ are the generating coordinates,
which are introduced to expand the relative motion
wavefunction. The delocalization parameter
$\epsilon(\mathbf{s}_i)$ is not an adjusted one but determined
variationally by the dynamics of the multi-quark system itself.
In this way, the multi-quark
system chooses its favorable configuration in the interacting process.

\subsection{Wave function}
In this paper, the resonating group method (RGM)~\cite{RGM}, a well-established method for studying a bound-state or a scattering problem, is used. The wave function of the four-quark system is of the form
\begin{equation}
\Psi = {\cal A } \left[[\psi^{L}\psi^{\sigma}]_{JM}\psi^{f}\psi^{c}\right].
\end{equation}
where $\psi^{L}$, $\psi^{\sigma}$, $\psi^{f}$, and $\psi^{c}$ are the orbital, the spin, the flavor and the color wave functions respectively, which are shown below. The symbol ${\cal A }$ is the anti-symmetrization operator. For the meson-meson structure, ${\cal A }$ is defined as
\begin{equation}
{\cal A } = 1-P_{13}-P_{24}+P_{13}P_{24}.
\end{equation}
where 1, 2 and 3, 4 represent the quarks in two meson clusters, respectively.

The orbital wave function is in the form of
\begin{equation}
\psi^{L} = {\psi}_{1}(\boldsymbol{R}_{1}){\psi}_{2}(\boldsymbol{R}_{2})\chi_{L}(\boldsymbol{R}).
\end{equation}
where $\boldsymbol{R}_{1}$ and $\boldsymbol{R}_{2}$ are the internal coordinates for the cluster 1 and cluster 2. $\boldsymbol{R} = \boldsymbol{R}_{1}-\boldsymbol{R}_{2}$ is the relative coordinate between the two clusters 1 and 2. The ${\psi}_{1}$ and ${\psi}_{2}$ are the internal cluster orbital wave functions of the clusters 1 and 2, and $\chi_{L}(\boldsymbol{R})$ is the relative motion wave function between two clusters, which is expanded by gaussian bases
\begin{eqnarray}
& & \chi_{L}(\boldsymbol{R}) = \frac{1}{\sqrt{4\pi}}(\frac{3}{2\pi b^2}) \sum_{i=1}^{n} C_{i}  \nonumber \\
&& ~~~~\times  \int \exp\left[-\frac{3}{4b^2}(\boldsymbol{R}-\boldsymbol{S}_{i})^{2}\right] Y_{LM}(\hat{\boldsymbol{S}_{i}})d\hat{\boldsymbol{S}_{i}}. ~~~~~
\end{eqnarray}
where $\boldsymbol{S}_{i}$ is called the generate coordinate, $n$ is the number of the gaussian bases, which is determined by the stability of the results. By doing this, the integro-differential equation of RGM can be reduced to an algebraic equation, generalized eigen-equation. Then we can obtain the energy of the system by solving this generalized eigen-equation. The details of solving the RGM equation can be found in Ref.~\cite{RGM}.
The flavor, the spin, and the color wave functions are constructed differently depending on different structures. Here, we investigate the meson-meson interaction, so wwe construct these wave functions within the meson-meson structure. As the first step, we give the wave functions of the meson cluster. The flavor wave functions of the meson cluster are shown below.
\begin{eqnarray}
\chi^{1}_{I_{11}} &=& u\bar{d},~~~~\chi^{2}_{I_{1-1}} = -d\bar{u},~~~~\chi^{3}_{I_{10}} = \sqrt{\frac{1}{2}}(d\bar{d}-u\bar{u}),\nonumber \\
\chi^{4}_{I_{\frac{1}{2}\frac{1}{2}}} &=& s\bar{d},~~~~\chi^{5}_{I_{\frac{1}{2}\frac{1}{2}}} = u\bar{s},~~~~\chi^{6}_{I_{00}} = s\bar{s}.
\end{eqnarray}
where the superscript of the $\chi$ is the index of the flavor wave function for a meson, and the subscript stands for the isospin  $I$ and the third component $I_{z}$. The spin wave functions of the meson cluster are:
\begin{eqnarray}
\chi^{1}_{\sigma_{11}} &=& \alpha\alpha,~~~~\chi^{2}_{\sigma_{10}} = \sqrt{\frac{1}{2}}(\alpha\beta+\beta\alpha), \nonumber \\
~~~~\chi^{3}_{\sigma_{1-1}} &=& \beta\beta,~~~~\chi^{4}_{\sigma_{00}} = \sqrt{\frac{1}{2}}(\alpha\beta-\beta\alpha).
\end{eqnarray}
and the color wave function of a meson is:
\begin{eqnarray}
\chi^{1}_{[111]} &=& \sqrt{\frac{1}{3}}(r\bar{r}+g\bar{g}+b\bar{b}).
\end{eqnarray}
Then, the wave functions for the four-quark system with the meson-meson structure can be obtained by coupling the wave functions of two meson clusters. Every part of wave functions are shown below. The flavor wave functions are:
\begin{eqnarray}
\psi^{f_{1}}_{22} &=& \chi^{1}_{I_{11}}\chi^{1}_{I_{11}},~~~~\psi^{f_{2}}_{\frac{3}{2}\frac{3}{2}} = \chi^{1}_{I_{11}}\chi^{5}_{I_{\frac{1}{2}\frac{1}{2}}},~~~~
\psi^{f_{3}}_{\frac{1}{2}\frac{1}{2}} = \chi^{6}_{I_{00}}\chi^{5}_{I_{\frac{1}{2}\frac{1}{2}}}
\nonumber\\
\psi^{f_{4}}_{11} &=& \sqrt{\frac{1}{2}}\left[\chi^{1}_{I_{11}}\chi^{3}_{I_{10}}-\chi^{3}_{I_{10}}\chi^{1}_{I_{11}}\right]
\nonumber\\
\psi^{f_{5}}_{00} &=& \sqrt{\frac{1}{3}}\left[\chi^{1}_{I_{11}}\chi^{2}_{I_{1-1}}-\chi^{3}_{I_{10}}\chi^{3}_{I_{10}}+\chi^{2}_{I_{1-1}}\chi^{1}_{I_{11}}\right].
\end{eqnarray}
The spin wave functions are:
\begin{eqnarray}
\psi^{\sigma_{1}}_{00} &=& \chi^{4}_{\sigma_{00}}\chi^{4}_{\sigma_{00}},\nonumber \\
\psi^{\sigma_{2}}_{00} &=& \sqrt{\frac{1}{3}}\left[\chi^{1}_{\sigma_{11}}\chi^{3}_{\sigma_{1-1}}-\chi^{2}_{\sigma_{10}}\chi^{2}_{\sigma_{10}}+\chi^{3}_{\sigma_{1-1}}\chi^{1}_{\sigma_{11}}\right],  \nonumber \\
\psi^{\sigma_{3}}_{11} &=& \chi^{4}_{\sigma_{00}}\chi^{1}_{\sigma_{11}},~~~~\psi^{\sigma_{4}}_{11} = \chi^{1}_{\sigma_{11}}\chi^{4}_{\sigma_{00}},  \nonumber \\
\psi^{\sigma_{5}}_{11} &=& \sqrt{\frac{1}{2}}\left[\chi^{1}_{\sigma_{11}}\chi^{2}_{\sigma_{10}}-\chi^{2}_{\sigma_{10}}\chi^{1}_{\sigma_{11}}\right],\nonumber \\
\psi^{\sigma_{6}}_{22} &=& \chi^{1}_{\sigma_{11}}\chi^{1}_{\sigma_{11}} .
\end{eqnarray}
The color wave function is:
\begin{eqnarray}
\psi^{c_{1}} &=& \chi^{1}_{[111]}\chi^{1}_{[111]}.
\end{eqnarray}
Finally, multiplying the wave functions $\psi^{L}$, $\psi^{\sigma}$, $\psi^{f}$, and $\psi^{c}$ according to the definite quantum number of the system, we can acquire the total wave functions.

\section{The result and discussion}
In this work, we investigate the interaction between two light mesons, which includes three types: two pseudoscalar mesons (PP), a pseudoscalar meson and a vector meson (PV), and two vector mesons (VV). As a preliminary calculation, only the $S-$wave systems are studied here, so we set the orbital angular momentum to zero. Due to the limit of the spin quantum number, the PP system has the total spin quantum number $S = 0$; the PV system has the total spin quantum number $S = 1$; while the VV system has three possible spin quantum numbers, which are $S = 0$, $1$, and $2$. Since the orbital angular momentum is $L=0$, the total  angular momentum can be $J = 0$ for PP systems, $J = 0$ and $1$ for PV systems, and $J = 0$, $1$, and $2$ for VV systems. Besides, the isospin of the four-quark systems with light quarks can be $I=0$, $\frac{1}{2}$, $1$, $\frac{3}{2}$, and $2$.
All possible channels for different systems are listed in Table~\ref{channels}.
\begin{table}[ht]
\caption{\label{kkk}Channels for different systems.}
\begin{tabular}{ccccc}
\hline \hline
I~J & ~~~~~~~Channel~~~~~~~~ \\ \hline
0~0~&$\pi\pi$,~$\eta\eta$,~$\eta\eta'$,~$\eta'\eta'$,~$K\bar{K}$,~$\phi\phi$,~$\omega\omega$,~$\omega\phi$,~$\rho\rho$,~$K^{*}\bar{K}^{*}$ \\
0~1~&$\pi\rho$,~$\eta\phi$,~$\eta\omega$,~$\eta'\phi$,~$\eta'\omega$,~$KK^{*}$,~$K\bar{K}^{*}$,~$\bar{K}K^{*}$,~$\omega\phi$,~$K^{*}K^{*}$,~$K^{*}\bar{K^{*}}$ \\
0~2~&$\phi\phi$,~$\omega\omega$,~$\rho\rho$,~$\omega\phi$,~$K^{*}\bar{K^{*}}$ \\
1~0~&$\pi\eta$,~$\pi\eta'$,~$KK$,~$K\bar{K}$,~$\omega\rho$,~$\phi\rho$,~$K^{*}K^{*}$,~$K^{*}\bar{K^{*}}$ \\
1~1~&$\pi\rho$,~$\pi\phi$,~$\pi\omega$,~$\eta\rho$,~$\eta'\rho$,~$KK^{*}$,~$K\bar{K}^{*}$,~$\bar{K}K^{*}$~,$\omega\rho$,~$\phi\rho$,~$\rho\rho$,~$K^{*}\bar{K^{*}}$ \\
1~2~&$\omega\rho$,~$\phi\rho$,~$K^{*}K^{*}$,~$K^{*}\bar{K^{*}}$ \\
2~0~&$\pi\pi$,~$\rho\rho$ \\
2~1~&$\pi\rho$ \\
2~2~&$\rho\rho$ \\
$\frac{1}{2}~0$~&$\pi K$,~$\pi \bar{K}$,$\eta K$,~$\eta\bar{K}$,~$\eta' K$,~$\eta' \bar{K}$,~$\phi K^{*}$,~$\phi \bar{K}^{*}$,~$\omega K^{*}$,\\
&~$\omega \bar{K}^{*}$,~$\rho K^{*}$,~$\rho \bar{K}^{*}$ \\
$\frac{1}{2}~1$~&$\pi K^{*}$,~$\pi \bar{K}^{*}$,~$\rho K$,~$\rho \bar{K}$,~$\eta K^{*}$,~$\eta \bar{K}^{*}$,~$\eta' K^{*}$,~$\eta' \bar{K}^{*}$,~$\omega K$,\\
&$\omega \bar{K}$,~$\phi K$,~$\phi \bar{K}$,~$\omega K^{*}$,~$\omega \bar{K}^{*}$,~$\phi K^{*}$,~$\phi \bar{K}^{*}$,~$\rho K^{*}$,~$\rho \bar{K}^{*}$\\
$\frac{1}{2}~2$~&$\phi K^{*}$,~$\phi\bar{K}^{*}$,~$\omega K^{*}$,~$\omega\bar{K}^{*}$,~$\rho K^{*}$,~$\rho \bar{K}^{*}$ \\
$\frac{3}{2}~0$~&$\pi K$,~$\pi\bar{K}$,~$\rho K^{*}$,~$\rho\bar{K^{*}}$ \\
$\frac{3}{2}~1$~&$\pi K^{*}$,~$\pi\bar{K}^{*}$,~$\rho K$,~$\rho\bar{K}$,~$\rho K^{*}$,~$\rho \bar{K}^{*}$ \\
$\frac{3}{2}~2$~&$\rho K^{*}$,~~$\rho\bar{K}^{*}$ \\
 \hline\hline
\end{tabular}
\label{channels}
\end{table}

To investigate the interaction between two mesons, we calculate the effective potentials of all the channels listed in Table~\ref{channels}. The effective
potential between two mesons is defined as $V(S) = E(S)-E(\infty)$, where $E(S)$ is the diagonal matrix element of
the Hamiltonian of the system in the generating coordinate. All the results are shown in Figs. 1-7, respectively.
Here, we note that some channels have the same effective potential. For example, the potentials of the $\pi K$ and $\pi \bar{K}$ are the same, because the contribution of each interaction term to the $\pi K$ and $\pi \bar{K}$ is the same. To save space, we only show the potential for one of them in the figure. And the same treatment is used in similar situations.

For the $I=0$ system, there are five PP channels and five VV channels for the system with $J=0$. It is clear that Fig. 1(a) shows the effective potential between PP mesons, while Fig. (b) shows the effective potential between VV mesons. We can see that the potential is repulsive for the $\eta\eta$, $\eta\eta'$, $\eta'\eta'$, and $K\bar{K}$ channels, while it is attractive for the $\pi\pi$ channel. In contrast, the potential is attractive for all channels of the VV systems. The attraction between $\rho$-$\rho$ is the largest one, followed by that of the $\omega\omega$ channel, which is larger than that of the $\omega\phi$ channel. Besides, the attraction of both the $\phi\phi$ and $K^{*}\bar{K}^{*}$ channels is very weak. For the $J=1$ system, there are seven PV channels and three VV channels. From Figs. 2(a) and (b) we can see that the potential for the $\eta' \phi$, $\eta\omega$ and $K\bar{K}^{*}$ channels is repulsive, while the one for other channels is attractive. In addition, the attraction of the VV channels is a little deeper than that of the PV channels. For the $J=2$ system, there are five VV channels. It is obvious in Fig. 2(c) that the there is a deep attraction between two $\rho$ mesons. The potential for all channels is attractive, except the $\phi\phi$ channels.

For the $I=1$ system, there are four PP channels and four VV channels for the system with $J=0$. From Fig. 3(a) we can see that the potential for the four VV channels is all attractive, while it is repulsive for the four PP channels. Moreover, there are four VV channels for the system with $J=2$, and the potential of all of them is attractive, which is shown in Fig. 3(b). For the system with $J=1$, there are seven PV channels and four VV channels, and the potential of them is shown in Fig. 4. The potential for the PV channels $\pi\omega$ and $\eta\rho$ is repulsive, while it is attractive for the PV channels $\pi\rho$ and $\eta'\rho$. Besides, there is very shallow attraction for the PV channels $\pi\phi$, $KK^{*}$ and $K\bar{K}^{*}$. By contrast, the potential for all VV channels is attractive, and that for both $\omega\rho$ and $\rho\rho$ channels is very deep.

For the $I=2$ system, it is obvious in Fig. 5 that the potential for both the $\pi\pi$ channel with $J=0$ and $\pi\rho$ channels with $J=1$ is repulsive, while the one for the $\rho\rho$ channel with $J=0$ and $J=2$ is attractive.

For the $I=\frac{1}{2}$ system, Fig. 6(a) shows the potential for three PP channels and three VV channels with $J=0$. One sees that the potential is repulsive for both two PP channels $\eta K$ and $\eta' K$, while it is attractive for the $\pi K$ channel. Moreover, it is obviously attractive for three VV channels. Fig. 6(b) shows the potential for six PV channels and three VV channels with $J=1$, from which we can see that the potential is repulsive for four PV channels $\eta K^{*}$, $\eta' K^{*}$, $\omega K$ and $\phi K$. However, the potential is attractive for other PV channels $\pi K^{*}$ and $\rho K$. Besides, it is also attractive for all VV channels. Meanwhile, Fig. 6(c) shows the potential for three VV channels with $J=2$. Obviously, the potential for both the $\omega K^{*}$ and $\rho K^{*}$ is attractive, while it is repulsive for the $\phi K^{*}$ channel.

For the $I=\frac{3}{2}$ system, Fig. 7 clearly shows that the potential is attractive for the VV channel $\rho\bar{K}^{*}$ with $J=0$, $1$ and $2$. However, there is no attractive potential for either the PP channel $\pi K$ with $J=0$ or the PV channels $\pi K^{*}$ and $\rho K$ with $J=1$.

From the above analysis, it is not difficult to find a rule that for most channels the interaction between two vector mesons is attractive; the one between a pseudoscalar meson and a vector meson is repulsive or weakly attractive; and the one between two pseudoscalar mesons is always repulsive. This law is similar with the one of the baryon-baryon interaction. In Ref.~\cite{PRC51}, the interaction between two decuplet baryons is almost deeply attractive; the one between a decuplet baryon and a octet baryon is always weakly attractive; and the one between two octet baryons is mostly repulsive or weakly attractive.
However, there is still some exception. For example, the potential for the PP channel $\pi\pi$ with $IJ=00$ is attractive, while it is repulsive for the $\pi\pi$ with $IJ=20$. This conclusion is consistent with most theoretical work. The $\pi\pi$ interaction has been studied as an important test of the strong interaction for a long time. Much attention has been paid to the isospin $I=0$ $\pi\pi$ S-wave interaction due to its direct relation to the $\sigma$ particle and the scalar glueball candidates ~\cite{PDG12,PM12,DV12,BA12,VV12,NN12,BSZ12}. Besides, the study of the $I=2$ $\pi\pi$ S-wave interaction is also necessary since a correct description of the $I=2$ $\pi\pi$ S-wave interaction is important for the extraction of the $I=0$ $\pi\pi$ S-wave interaction from experimental data~\cite{ZouBS}. Therefore, the study of the $\pi\pi$ scattering process and exploring the resonance states is our further work. Moreover, the $S-$wave $\phi\phi$ channel is a special state, which is composed of two vector mesons but with the repulsive interaction. So it is difficult to use the $S-$wave $\phi\phi$ state to explain the $X(2239)$
at present calculation. To explore the candidate of $X(2239)$, the study of the high partial wave of the $\phi\phi$ state will be performed in future work.
We will explain why the interaction is repulsive between two $\phi$ mesons later.

In addition, the deep attraction between two vector mesons also attracts great attention to the systems composed of two vector mesons.
For the $S-$wave $\rho\rho$ state, the deepest effective attraction of the states with different quantum numbers is from about $100$ MeV to $200$ MeV.
We find that the attraction of the state with isospin $I=0$ is larger than that with $I=2$. For the $I=0$ $\rho\rho$,
the attraction of the state with the angular momentum $J=0$ is larger than that with $J=2$; while for the $I=2$ $\rho\rho$, the attraction of the state with $J=0$ is smaller than that with $J=2$.
Nevertheless, there is great attraction in the $\rho\rho$ state, which makes it more possible to form some bound states or resonance states.

\begin{figure}[ht]
\begin{center}
\epsfxsize=3.0in \epsfbox{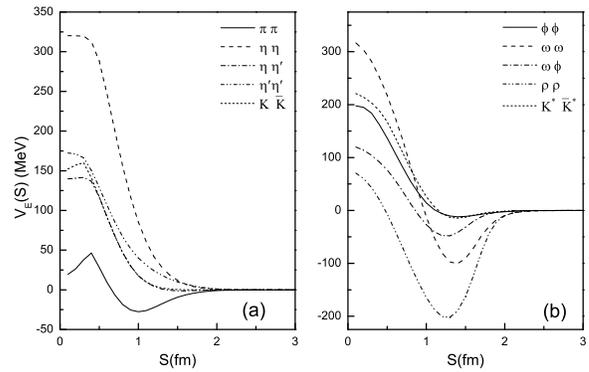} \vspace{-0.1in}
\caption{The Figures (a) and (b) are the effective potentials of all channels for the system with $IJ=00$.}
\end{center}
\end{figure}

\begin{figure}[ht]
\begin{center}
\epsfxsize=3.2in \epsfbox{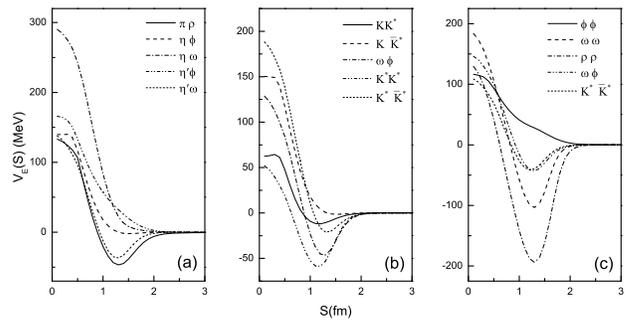} \vspace{-0.1in}
\caption{The Figures (a) and (b) are the effective potentials of all channels for the system with $IJ=01$ and the Figure (c) is for the system with $IJ=02$.}
\end{center}
\end{figure}

\begin{figure}[ht]
\begin{center}
\epsfxsize=3.0in \epsfbox{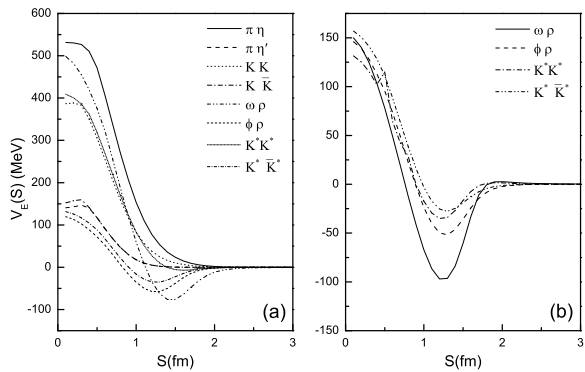} \vspace{-0.1in}
\caption{The Figure (a) is the effective potentials of all channels for the system with $IJ=10$ and the Figure (b) is for the system with $IJ=12$.}
\end{center}
\end{figure}

\begin{figure}[ht]
\begin{center}
\epsfxsize=3.0in \epsfbox{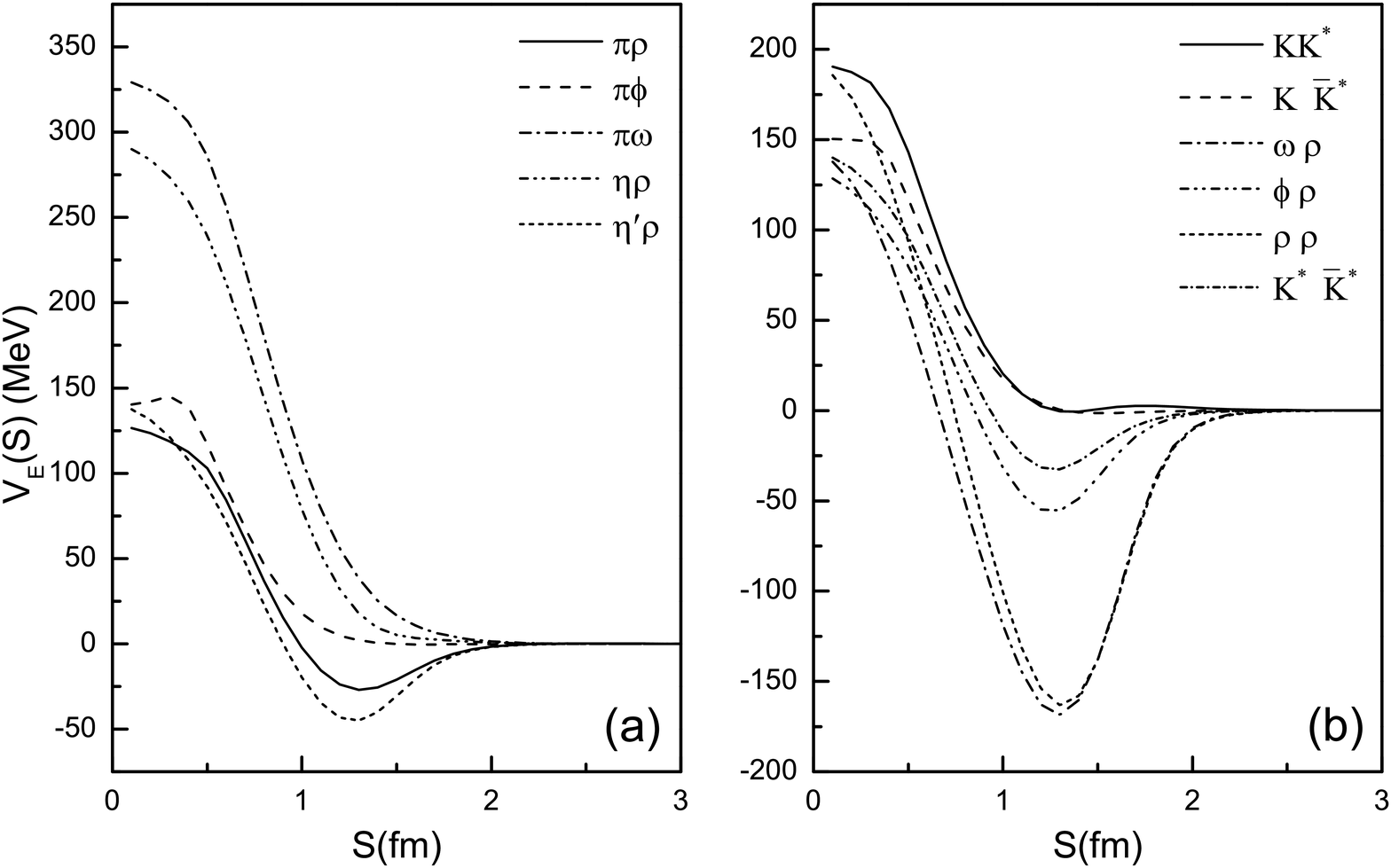} \vspace{-0.1in}
\caption{The Figure (a) and (b) are the effective potentials of all channels for the system with $IJ=11$.}
\end{center}
\end{figure}

\begin{figure}[ht]
\begin{center}
\epsfxsize=3.2in \epsfbox{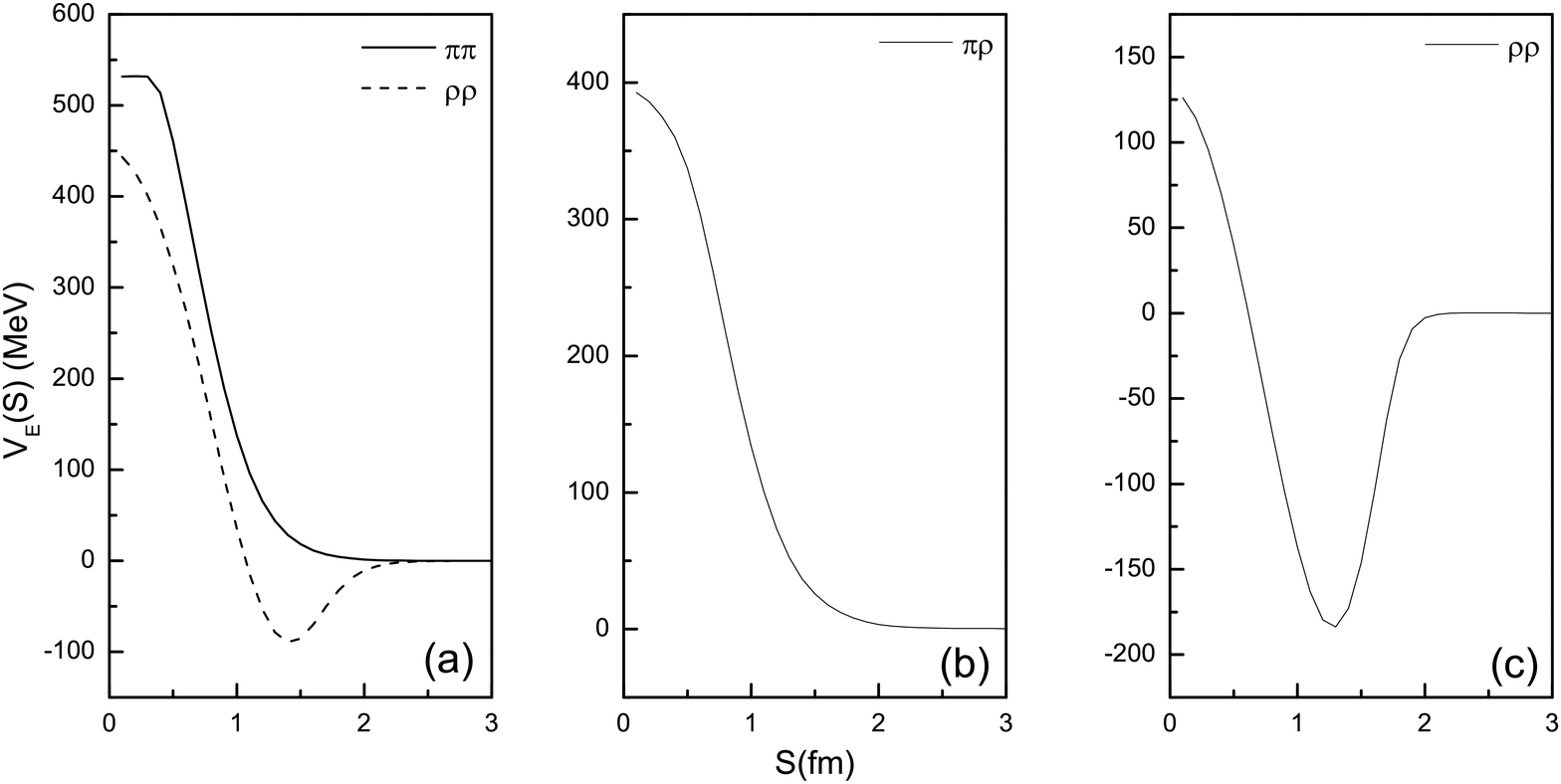} \vspace{-0.1in}
\caption{The Figure (a) is the effective potentials of all channels for the system with $IJ=20$, Figure (b) is for the system with $IJ=21$ and Figure (c) is for the system with $IJ=22$.}
\end{center}
\end{figure}

\begin{figure}[ht]
\begin{center}
\epsfxsize=3.0in \epsfbox{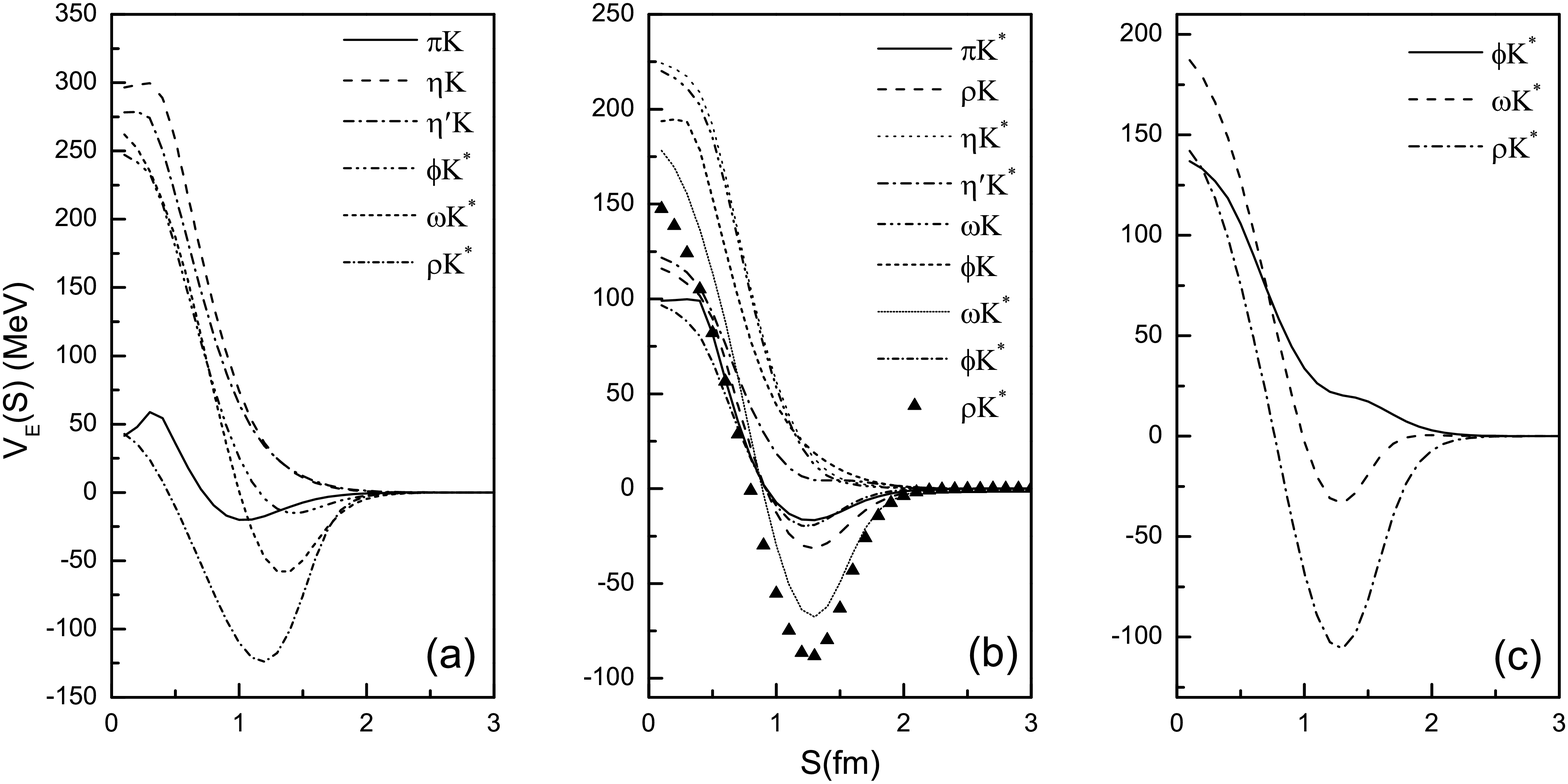} \vspace{-0.1in}
\caption{The Figure (a) is the effective potentials of all channels for the system with $IJ=\frac{1}{2} 0$, Figure (b) is for the system with $IJ=\frac{1}{2} 1$ and Figure (c) is for the system with $IJ=\frac{1}{2} 2$.}
\end{center}
\end{figure}

\begin{figure}[ht]
\begin{center}
\epsfxsize=3.2in \epsfbox{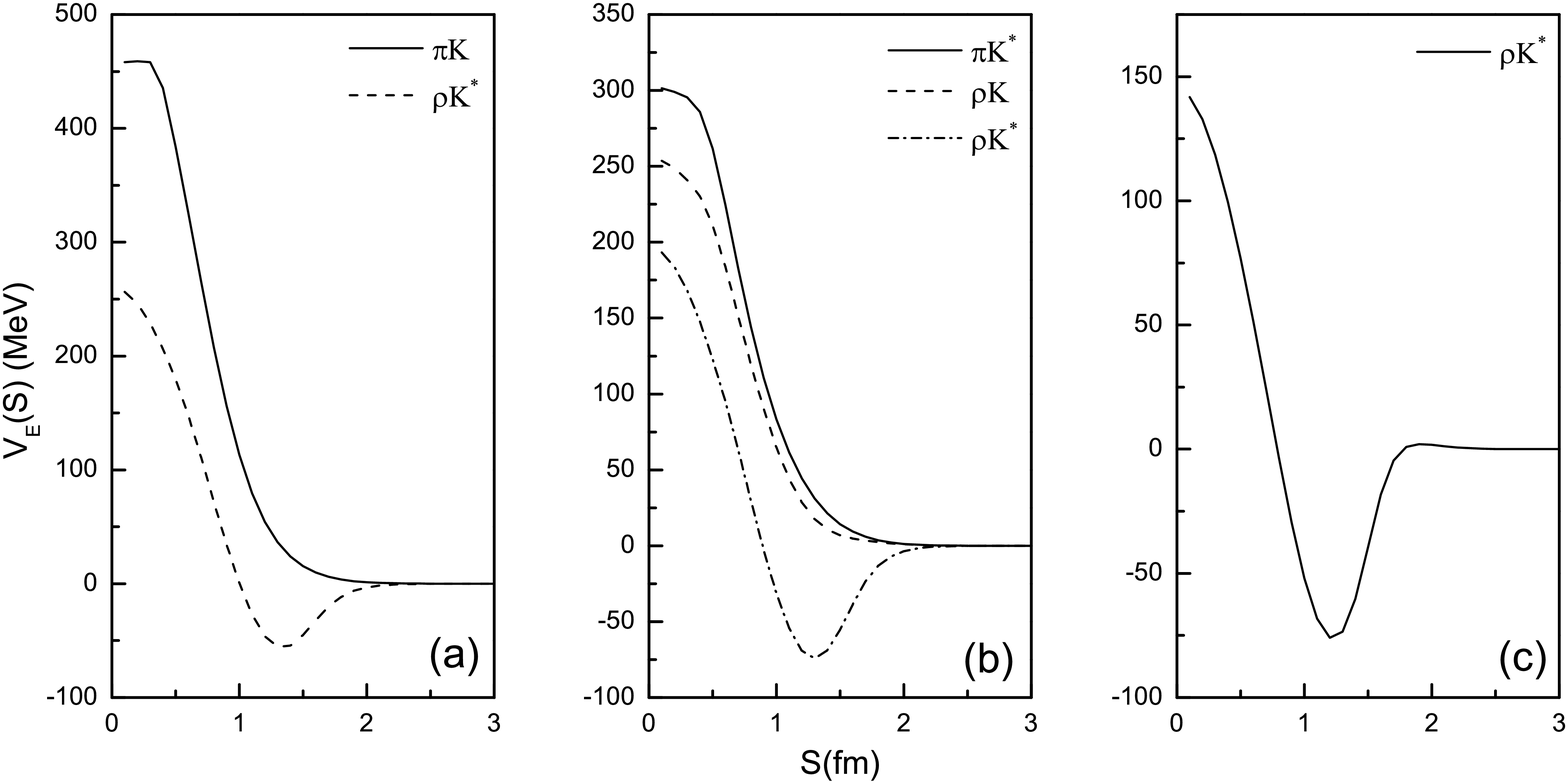} \vspace{-0.1in}
\caption{The Figure (a) is the effective potentials of all channels for the system with $IJ=\frac{3}{2} 0$, Figure (b) is for the system with $IJ=\frac{3}{2} 1$ and Figure (c) is for the system with $IJ=\frac{3}{2} 2$.}
\end{center}
\end{figure}

In order to explore the contribution of each interaction term to the system, the interaction from various terms, the kinetic energy $(V_{\nu k})$, the confinement $(V_{con})$, the one-gluon exchange $(V_{oge})$, and the one-boson exchange $(V_{\pi},V_{K},V_{\eta})$ are studied. To save space, we take a few states for examples, which are the $\eta\rho$ and $\rho\rho$ states with $IJ=11$, and the $\pi\pi$ and $\rho\rho$ states with $IJ=20$. The contribution of each interaction term is shown in Figs. 8 and 9. From the Fig. 8 we can see that the kinetic energy term provides an attractive interaction for both $IJ=11$ $\eta\rho$ and $\rho\rho$ states, and this attraction of the $\rho\rho$ state is obviously larger than that of the $\eta\rho$ state. Besides, the one-gluon exchange interaction provides deep attraction for the $\rho\rho$ state, while it is repulsive for the $\eta\rho$ state. In contrast, the contribution of other items is small. So both the kinetic energy and the one-gluon exchange interactions provide attraction, which leads a large total attraction for the $\rho\rho$ state; while for the $\eta\rho$, the repulsive interaction by the one-gluon exchange and the attractive interaction by the kinetic energy almost cancel each other out, which result in the total repulsive interaction. Towards the $\pi\pi$ and $\rho\rho$ states with $IJ=20$, we find that the one-gluon exchange provides repulsive interaction for both $\pi\pi$ and $\rho\rho$ states. However, the kinetic energy interaction is attractive for the $\rho\rho$ state, which leads to a total attraction; while it is repulsive for the $\pi\pi$ state, which leads to a total repulsive interaction. From the above discussion, we can see that the kinetic energy interaction plays an important role in providing attractions, which relates to the intermediate-range attraction mechanism in QDCSM.

\begin{figure}[ht]
\begin{center}
\epsfxsize=3.0in \epsfbox{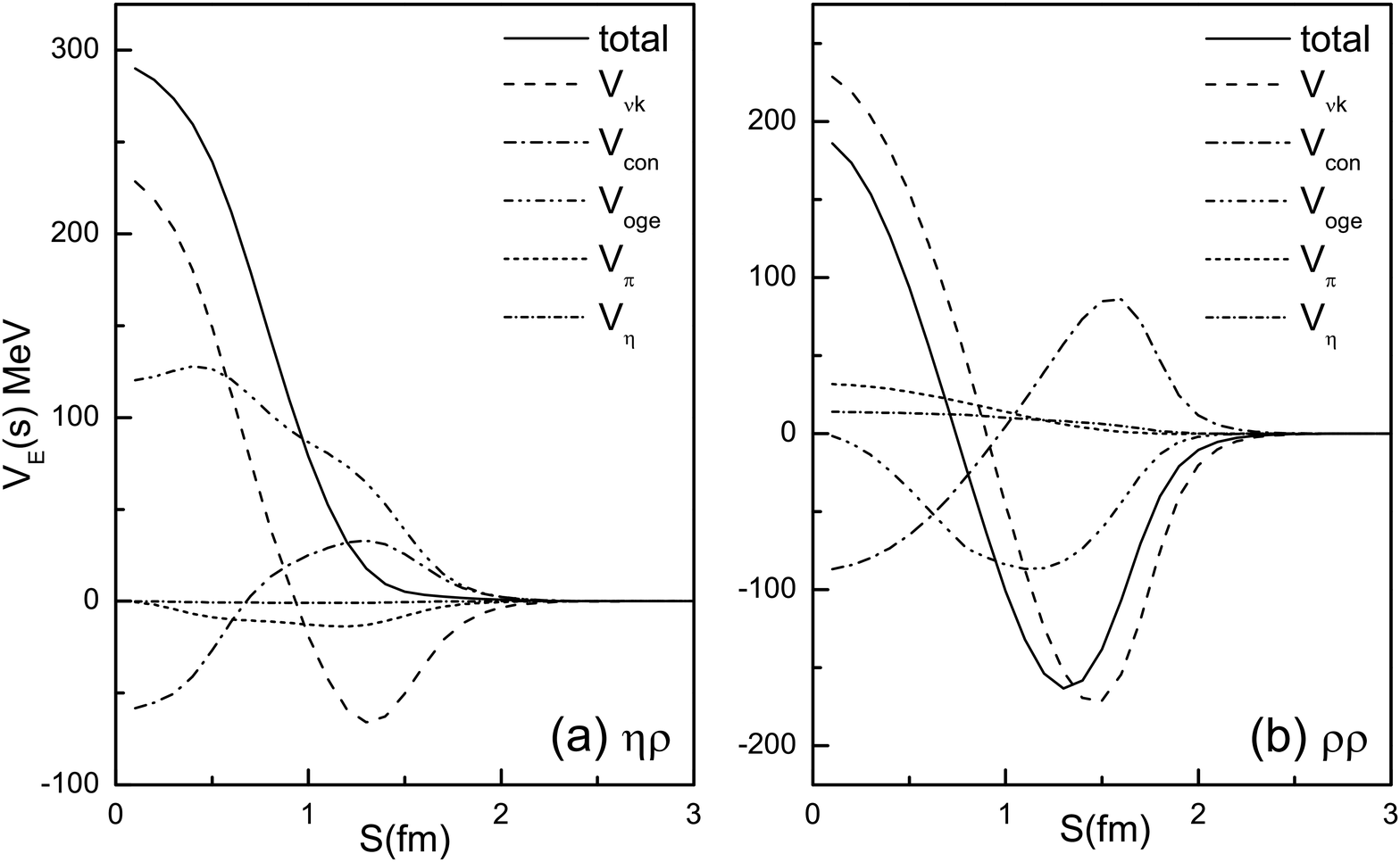} \vspace{-0.1in}
\caption{The contributions to the effective potentials from various terms of $\eta\rho$ and $\rho\rho$ with $ IJ=11$.}
\end{center}
\end{figure}

\begin{figure}[ht]
\begin{center}
\epsfxsize=3.0in \epsfbox{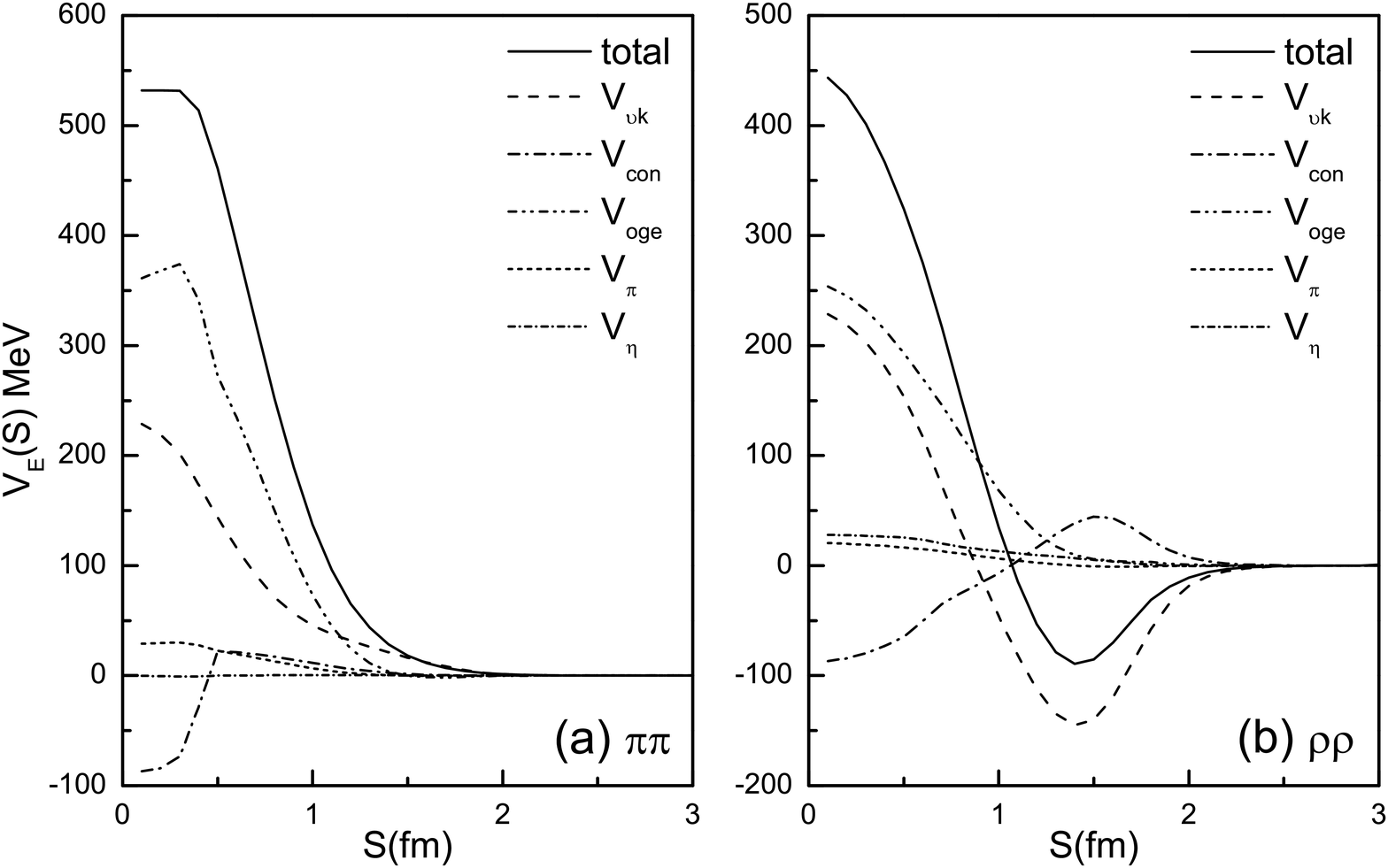} \vspace{-0.1in}
\caption{The contributions to the effective potentials from various terms of $\pi\pi$ and $\rho\rho$ with $ IJ=20$.}
\end{center}
\end{figure}

In QDCSM, two ingredients were introduced: quark delocalization and color screening, the former is to enlarge the model variational space to
take into account the mutual distortion or the internal excitations of nucleons in the course of interaction, and the latter is assuming that
the quark-quark interaction dependents on quark states and aims to take into account the QCD effect which has not been included in the two body confinement and effective one gluon exchange yet. In this model, the intermediate-range attraction is achieved by the quark delocalization, which is like the electron delocalization in
molecules. The color screening is needed to make the quark delocalization effective. It is worth noting that the delocalization parameter is not an adjusted one but determined
variationally by the dynamics of the system itself. Here, we show the variation of the delocalization parameter, which relates to the intermediate-range attraction for the different states.

\begin{figure}[ht]
\begin{center}
\epsfxsize=3.0in \epsfbox{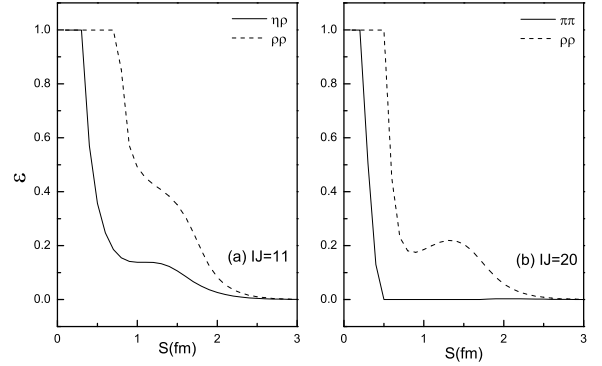} \vspace{-0.1in}
\caption{The delocalization parameter $\varepsilon$ of $\eta\rho$ and $\rho\rho$ with $IJ=11$ and $\pi\pi$ and $\rho\rho$ with $IJ=20$.}
\end{center}
\end{figure}

In Fig. 10(a), the delocalization parameter of the $\rho\rho$ state with $IJ=11$ is close to $1$ when the distance $S$ between two mesons is less than $0.8$ fm, which means that the quarks are willing to run between different clusters, thereby reducing the kinetic energy and introducing the attractive interaction. In contrast, the delocalization parameter of the $\eta\rho$ state with $IJ=11$ approaches to $1$ when $S \leq 0.3$ fm, and it quickly approaches to $0$ as the distance increases. Although it can also reduce the kinetic energy, the reduction of the $\eta\rho$ state is not as great as that of the $\rho\rho$ state, so the attraction of the $\eta\rho$ state is smaller than that of the $\rho\rho$ state as shown in Fig. 8. The case is similar for the $\pi\pi$ and $\rho\rho$ states with $ IJ=20$. From Fig. 10(b) we can see that the delocalization parameter of the $\rho\rho$ state with $IJ=20$ is close to $1$ when the distance $S \leq 0.5$ fm, then it approaches to $0$ more quickly than that of the $\rho\rho$ state with $IJ=11$, that is why the kinetic energy of the $\rho\rho$ state with $IJ=11$ is lower than that of the $\rho\rho$ state with $IJ=20$. With regard to the $\pi\pi$ state with $IJ=20$, the delocalization parameter is close to $1$ only when the distance $S \leq 0.2$ fm, and it approaches to $0$ more quickly than that of the $\eta\rho$ state with $IJ=11$, so the kinetic energy of $\pi\pi$ state with $IJ=20$ is much higher than that of the $\eta\rho$ state with $IJ=11$, even it cannot provide the attractive interaction.

The variation of the delocalization parameter can also be used to explain the repulsive interaction of the $\phi\phi$ state. We show the variation of the delocalization parameter for the $IJ=02$ system in Fig. 11.
Comparing the $\phi\phi$ and $\rho\rho$ states, although both of them are composed of two vector mesons, the quark component of the $\phi\phi$ state is $s\bar{s}s\bar{s}$, and the one of the $\rho\rho$ state is $q\bar{q}q\bar{q}$ ($q=u$ or $d$). The mass of the strange quark is heavier than that of the nonstrange quark, so the strange quark is less willing to run between two clusters. As shown in Fig. 11, the delocalization parameter of the $\phi\phi$ state is close to $1$ only when the distance $S \leq 0.2$ fm, and it approaches to $0$ very quickly, so it cannot reduce the kinetic energy too much. So the interaction between two $\phi$ is repulsive. Besides, by comparing Fig. 11 and Fig. 2(c), we find that the variation of the delocalization parameter relates to the interaction between two mesons. The more the quark runs between the two mesons, the greater attraction between the two mesons has.

\begin{figure}[ht]
\begin{center}
\epsfxsize=3.0in \epsfbox{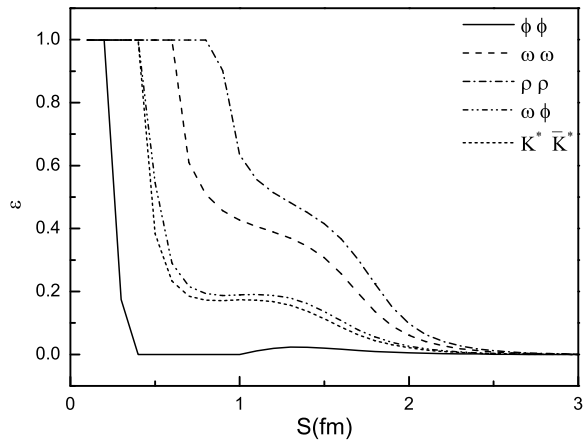} \vspace{-0.1in}
\caption{The delocalization parameter $\varepsilon$ for the $IJ=02$ system.}
\end{center}
\end{figure}

\section{Summary}
The study of the hadron-hadron interaction is one of the critical issues in the hadron physics. In this work, we investigate the meson-meson interaction in the fully light four-quark system in the framework of the QDCSM. The effective potentials of all the $S-$wave states,
the contribution of each interaction term to the states, as well as the variation of the delocalization parameter, which relates to the intermediate-range attraction for the different states are all studied in this work.

Our results show that for most states the interaction between two vector mesons is attractive; the one between a pseudoscalar meson and a vector meson is repulsive or weakly attractive; and the one between two pseudoscalar mesons is always repulsive. However, there is still some exception. The interaction of the $IJ=00$ $\pi\pi$ channel is attractive, while the one of the $IJ=02$ $\phi\phi$ channel is repulsive. This law is similar with the one of the baryon-baryon interaction. In the dibaryon systems, the interaction between two $\Delta$ baryons is deeply attractive, which leads to the well-known bound state $d^{*}$. So we should pay more attention to the four-quark system composed of two vector mesons here. Among all these states, the $S-$wave $\rho\rho$ state, especially the state with $IJ=00$ is more likely to be a bound state or a resonance state. We will continue to study these states in further work.

The study of the contribution of each interaction term shows that both the one-gluon exchange and the kinetic energy interaction play an important role in the interaction between two mesons. Besides, the kinetic energy relates to the intermediate-range attraction mechanism in QDCSM, which is achieved by the quark delocalization. The delocalization parameter approaching to $1$ means that the quarks are more willing to run between the two mesons, thereby reduce the kinetic energy and introduce the attractive interaction. Our results show that the more the quark runs between the two mesons, the greater attraction between the two mesons has.

\acknowledgments{This work is supported partly by the National Science Foundation
of China under Contract Nos. 11675080, 11775118 and 11535005.}

\end{document}